\documentclass[11pt,fleqn]{article}

\usepackage{cite,latexsym,amssymb,amsfonts}
\usepackage{graphicx}
\usepackage{epstopdf}

\def\noi{\noindent}

\newcommand{\Title}[1]{\noi {{\Large\bf #1}}\\[1ex]}

\def\Aunames#1{\noi{\bf #1}}
\def\auth#1{${}^{#1}$}
\def\Addresses#1{\medskip\noi \protect
	\begin{description}\itemsep -3pt {\it #1} \end{description}}
\def\addr#1#2{\item[${}^{#1}$]{\it #2}}

\newcommand{\Abstract}[1]{\vskip 2mm \begin{center}
        \parbox{16.4cm}{\small\noi #1} \end{center}\medskip}

\def\email#1#2{\footnotetext[#1]{e-mail: #2}\addtocounter{footnote}{1}}

\def\nqq{\hspace*{-2em}}
\def\nhq{\hspace*{-0.5em}}

\def\cm{\hspace*{1cm}}
\def\inch{\hspace*{1in}}

\def\ten#1{\mbox{$\cdot 10^{#1}$}}

\def\Jl#1#2{#1 {\bf #2},\ }

\def\ApJ#1 {\Jl{Astroph. J.}{#1}}
\def\CQG#1 {\Jl{Class. Quantum Grav.}{#1}}
\def\DAN#1 {\Jl{Dokl. AN SSSR}{#1}}
\def\GC#1 {\Jl{Grav. Cosmol.}{#1}}
\def\GRG#1 {\Jl{Gen. Rel. Grav.}{#1}}
\def\JETF#1 {\Jl{Zh. Eksp. Teor. Fiz.}{#1}}
\def\JETP#1 {\Jl{Sov. Phys. JETP}{#1}}
\def\JHEP#1 {\Jl{JHEP}{#1}}
\def\JMP#1 {\Jl{J. Math. Phys.}{#1}}
\def\NPB#1 {\Jl{Nucl. Phys. B}{#1}}
\def\NP#1 {\Jl{Nucl. Phys.}{#1}}
\def\PLA#1 {\Jl{Phys. Lett. A}{#1}}
\def\PLB#1 {\Jl{Phys. Lett. B}{#1}}
\def\PRD#1 {\Jl{Phys. Rev. D}{#1}}
\def\PRL#1 {\Jl{Phys. Rev. Lett.}{#1}}

\def\al{&\nhq}
\def\lal{&&\nqq {}}
\def\eq{Eq.\,}
\def\eqs{Eqs.\,}
\def\beq{\begin{equation}}
\def\eeq{\end{equation}}
\def\bear{\begin{eqnarray}}
\def\bearr{\begin{eqnarray} \lal}
\def\ear{\end{eqnarray}}
\def\earn{\nonumber \end{eqnarray}}

\def\nnn{\nonumber\\ \lal }

\def\yy{\\[5pt] {}}
\def\yyy{\\[5pt] \lal }
\def\eql{\al =\al}

\def\diag{\mathop{\rm diag}\nolimits}

\def\const{{\rm const}}

\begin{document}
\twocolumn[

\Title{Wormholes without exotic matter in Einstein-Cartan theory}

\Aunames{K. A. Bronnikov\auth {a,b,c,1} and A. M. Galiakhmetov\auth{d,2}}

\Addresses{
\addr a {VNIIMS, Ozyornaya ul. 46, Moscow 119361, Russia}
\addr b	{Peoples' Friendship University of Russia,
	ul. Miklukho-Maklaya 6, Moscow 117198, Russia}
\addr c	{National Research Nuclear University ``MEPhI''
	(Moscow Engineering Physics Institute), Moscow, Russia}
\addr d	{Donetsk National Technical University, ul. Kirova 51, 84646,
	Gorlovka, Ukraine}
	}

\Abstract {We study the possible existence of static traversable wormholes
     without invoking exotic matter in the framework of the Einstein--Cartan
     theory. A family of exact static, spherically symmetric wormhole
     solutions with an arbitrary throat radius, with flat or AdS asymptotic
     behavior, has been obtained with sources in the form of two
     noninteracting scalar fields with nonzero potentials. Both scalar
     fields are canonical (that is, satisfy the weak energy condition), one
     is minimally and the other nonminimally coupled to gravity, and the
     latter is a source of torsion.
     }

%

\bigskip
\bigskip

] 
\email 1 {kb20@yandex.ru}
\email 2 {agal17@mail.ru}

\section{Introduction}

  Recent cosmic observations favor an isotropic, spatially flat Universe,
  which is at present expanding with acceleration. Establishment of the
  origin of this acceleration has become one of the most important problems
  in cosmology and even in theoretical physics as a whole. Different
  theoretical models trying to explain it have been put forward (see, e.g.,
  the reviews [1--4] and references therein). In the
  framework of general relativity (GR), this expansion may be ascribed to a
  source in the form of an unknown substance with a large negative pressure,
  called dark energy (DE), and its most popular model agreeing with
  observations is the cosmological constant. Many models have been suggested
  in alternative theories of gravity, in particular, in the Einstein--Cartan
  theory (ECT), see, e.g., [5--7]. This theory [8--11] is an extension of GR 
  to a space-time with torsion, and it reduces to GR if the torsion vanishes.

  The ECT is the simplest version of the Poincar\'e gauge theory of gravity
  (PGTG), in which the torsion is not dynamic because the gravitational
  action is proportional to the curvature scalar of Riemann--Cartan
  space--time. In this sense, the ECT is a degenerate gauge theory
  [11--14]. This shortcoming is absent in the full PGTG
  since its gravitational Lagrangian includes invariants quadratic in the
  curvature and torsion tensors. Nevertheless, the ECT is a viable theory
  of gravity whose observational predictions are in agreement with the
  classical tests of GR, and it significantly differs from GR only at very
  high densities of matter \cite{HO07,Trautman,BH11}.

  The interest in the ECT has recently grown in connection with the fact
  that torsion arises naturally in supergravity [17--19], Kaluza--Klein 
  [20--22] and superstring [23--25] theories. $f(R)$ gravity with torsion
  has been developed \cite{CCSV07,CCSV08} as one of the simplest extensions
  of the ECT. In \cite{CCSV08} it has been demonstrated that, in $f(R)$
  gravity, torsion can be a geometric source of accelerated expansion.

  Some cosmological models in ECT have turned out to be nonsingular, with a
  cosmological singularity replaced by a bounce \cite{G11}, while a similar
  result is achieved in GR only by invoking ``exotic'' sources, violating
  the energy conditions respected by usual matter, above all, the weak and
  null energy conditions (WEC and NEC). It is well known that in GR such a
  violation is a necessary condition for the existence of traversable
  wormholes \cite{hoh-vis}. Wormholes as possible time machines, or
  ``tunnels'' between universes or distant parts of the same universe, are a
  subject of particular interest, for reviews see \cite{vis-book, lobo-rev,
  BR-book} and references therein.

  Well known are wormhole solutions of GR with minimally coupled phantom
  scalar fields (those with a wrong sign of kinetic energy), with or without
  electric or magnetic charges as well as potentials, see [32--35]
  and references therein, while with normal
  fields having positive kinetic energy wormholes are impossible. For
  scalar fields with nonminimal couplings the situation is more subtle:
  even with normal fields there exist some special wormhole solutions (see
  \cite{br73, BVis99} for solutions with conformal coupling and
  \cite{BVis00, br-96, br-gr02} for other couplings). However, in all such
  cases the wormhole solutions inevitably contain regions where the
  effective gravitational constant becomes negative, that is, the
  gravitational field itself becomes a phantom \cite{br-JMP,br-star07}.
  Moreover, all such configurations, whose existence is connected with the
  phenomenon of conformal continuation \cite{br-Pol,br-JMP}, turn out to be
  unstable under radial perturbations \cite{br-gr01,br-gr02,br-gr04}.

  An aspect of interest for scalar fields with nonminimal coupling is that
  the existence of a throat as a local property of space-time does not yet
  mean that the configuration as a whole is a wormhole. It has been shown,
  in particular, that in the Brans-Dicke scalar-tensor theory of gravity,
  throats in static, spherically symmetric solutions can exist with any
  value of the coupling constant $\omega$ whereas wormholes as global
  entities only exist in the ``phantom range'' $\omega < -3/2$ \cite{BSS10}.

  One more possible source of unusual geometries without WEC or NEC
  violation is a vortex gravitational field \cite{krechet1, krechet2} existing
  in rotating configurations. Wormholes have been found among rotating
  cylindrically symmetric configurations \cite{BKS12} without exotic matter,
  but the main problem with them is their non-flat asymptotic behavior which
  does not allow them to be interpreted as objects in the observable
  Universe. Torsion in the ECT is to a certain extent similar to rotation
  in GR, therefore it is natural to expect that wormholes can exist in it
  without WEC or NEC violation.

  We here seek regular static, spherically symmetric configurations with
  normal, non-phantom scalar fields in the framework of the ECT and find
  some examples of wormhole solutions involving a nonminimally coupled
  scalar field as a source of torsion.

  The paper is organized as follows. In Section 2 we present the ECT
  equations both in the general case and for static, spherically symmetric
  configurations involving two scalar fields of which one is minimally and
  the other nonminimally coupled with space-time curvature. Section 3 is
  devoted to finding and analyzing the properties of a family of exact
  solutions, and Section 4 is a discussion.

\section{Field equations}

  We start with the action
\bearr      \label{1}
        S = \int \sqrt{- g}d^{4}x \Bigl [- \frac{R}{2\kappa }+
        \frac{\eta_1}{2} \Bigl (\phi_{,k}\phi^{,k}
        + \xi R\phi^2 \Bigr )
\nnn \inch
        - V(\phi ) +  \frac{\eta_2}{2} \psi_{,k}\psi^{,k}
	- W(\psi )\Bigr ],
\ear
  where $R [\Gamma]$ is the curvature scalar obtained from the full
  connection $\Gamma^{k}_{ij} = \{^{k}_{ij}\} + S_{ij\cdot }^{\ \ {k}}
  + S^k_{\cdot {ij}} + S^k_{\cdot {ji}}$; here $\{^{k}_{ij}\}$ are
  Christoffel symbols of the second kind for the metric $g_{ik}$;
  $S_{ij\cdot }^{\ \ {k}} =\Gamma^{k}_{[ij]}$ is the torsion tensor;
  $\kappa = 8\pi G$, $G$ being the Newtonian gravitational constant;
  $\phi $ and $\psi $ are two noninteracting scalar fields with the
  potentials $V(\phi)$ and $W(\psi)$, respectively. The constants
  $\eta_{1,2} = \pm 1$ correspond to either usual, canonical ($\eta = + 1$)
  or phantom ($\eta = -1$) scalar fields.

  The metric $g_{ik}$ has the signature ($+\ -\ -\ -$), the Riemann
  and Ricci tensors are defined as
\[
        R^{\ \ \ m}_{ijk\cdot } = \Gamma^{m}_{jk,i} - \Gamma^{m}_{ik,j}
        + \Gamma^{m}_{ip}\Gamma^{p}_{jk} - \Gamma^{m}_{jp}\Gamma^{p}_{ik}
\]
  and $R_{jk} = R^{\ \ \ i}_{ijk\cdot }$. We should note that, in the
  framework of ECT, a scalar field nonminimally coupled to gravity
  gives rise to torsion, even though the scalar field has zero spin.
  It follows from (\ref {1}) that the torsion can interact with a
  scalar field only through its trace: $S_{i} = S_{ik\cdot }^{\ \ k}$
  (see \cite{KS97}). Hence, the curvature scalar $R(\Gamma) =
  g^{jk}R_{jk}$ can be presented in the form \cite{KS97}
\beq\label2
         R [\Gamma] = R[\{ \}] + 4\nabla_k S^k  -(8/3)S_k S^k  \ ,
\eeq
  where $R [\{ \}]$ is the Riemannian part of the curvature built from
  the Christoffel symbols; $\nabla_k $ is the covariant derivative of
  Riemannian space.

  Varying the action with the Lagrangian (\ref {1}) in $g_{ij}$,
  $S_k$, $\phi$ and $\psi$, we obtain the following set of equations:
\bear             \label{3}
  	G_{ij}[\{\}] \eql \kappa (T_{ij}[\phi] + T_{ij}[\psi])
			+ \Lambda_{ij},
\yy   		\label{4}
  	S^k \eql \frac{3}2  \xi \Psi \phi \phi^{,k},
\yy		\label{5}
  	\Box \phi \lal - \xi \phi R [\Gamma] + \eta_1 dV/d\phi = 0,
\yy   		\label{6}
  	\Box \psi \lal + \eta_2 dW/d\psi = 0,
\ear
  where
\bearr         \label{7}
  	T_{ij}[\phi] = \eta_1 \biggl \{\phi_{,i}\phi_{,j}
                 - \frac 12 \Big[\phi_{,m}\phi^{,m} + \xi R[\{ \}]\phi^2
\nnn \ \ \
          - 2\eta_1 V(\phi )\Big]\, g_{ij}
                 + \xi \Big[-4S_{(i}\nabla_{j)} + 2g_{ij}S^{n}\nabla_{n}
\nnn \ \ \
          - \nabla_{i}\nabla_{j} + g_{ij}\Box + R_{ij}[\{ \}] -
               \Lambda_{ij}\Big]\phi^2 \biggr\},
\yyy               \label{8}
  	T_{ij}[\psi] = \eta_2 \Bigl (\psi_{,i}\psi_{,j}
               - \frac 1 2 \psi_{,m}\psi^{,m}g_{ij}\Bigr) + W(\psi )g_{ij},
\yyy		     \label{9}
  	\Lambda_{ij} = \frac{8}{3} S_{i}S_{j} - \frac{4}{3} S_k S^k g_{ij}.
\ear
  Here $\Box$ is the d'Alembertian operator of Riemannian space, and we
  denote $\Psi =\kappa (\eta_1 - \xi \kappa \phi^2 )^{- 1}$.

  It is not difficult to verify that the effective scalar-torsion
  stress-energy tensor $T^{\rm (eff)}_{ij}[\phi]$
\beq\label{10}
        T^{\rm (eff)}_{ij} [\phi] = T_{ij}[\phi] + \kappa^{- 1}\Lambda_{ij},
\eeq
  and the scalar stress-energy tensor $T_{ij}[\psi]$ are separately
  covariantly conserved since no explicit coupling is assumed between the
  scalar fields:
\beq\label{11}
         \nabla^j T^{\rm (eff)}_{ij} [\phi] = \nabla^j T_{ij}[\psi] = 0.
\eeq

  The general static, spherically symmetric metric can be written in
  the form
\beq\label{12}
   	ds^2  = A(u)dt^2  - \frac{du^2}{A(u)} - r^2 (u)d\Omega^2
\eeq
  in terms of the so-called quasiglobal radial coordinate \cite{BR-book},
  where $g_{00} = A(u)$ may be called the redshift function while $r(u)$ is
  the area function; $d\Omega^2  = d\theta^2  + \sin^2 \theta d\varphi^2$
  is the linear element on a unit sphere. (As usual, the metric is only
  formally static: it is really static if $A > 0$, but it describes a
  Kantowski--Sachs (KS) type cosmology if $A < 0$, and $u$ is then a temporal
  coordinate.) We also consider $\phi = \phi(u)$ and $\psi = \psi(u)$.

  The nonvanishing components of the effective scalar-torsion
  stress-energy tensor are given by
\bearr             \label{13}
   	T^{1\ \rm (eff)}_1 [\phi] = \eta_1 \xi \phi^2 G^1_1 [\{\}] + Y
   	+ \eta_1 \Big[ 2\xi \phi \phi''A
\nnn \ \
   	+ \xi \phi \phi'A'+ (- 1 + 2\xi + 6\xi^2 \phi^2 \Psi)\phi'^2 A \Big],
\yyy                \label{14}
   	T^{2\ \rm (eff)}_2 [\phi] = T^{3 \rm eff}_3 [\phi]
\nnn \ \
   	= \eta_1 \xi \phi^2 G^2_2 [\{\}] + Y
   		+ 2\eta_1 \xi \frac{r'}{r}\phi \phi'A,
\yyy                      \label{15}
   	T^{0 \rm eff}_0 [\phi] = \eta_1 \xi \phi^2 G^0_0[\{\}] + Y
   		+ \eta_1 \xi \phi \phi'A',
\ear
  where the prime means $d/du$ and
\bearr             \label{16}
   	Y = V(\phi ) + \eta_1 A \biggl[ -2\xi \phi \phi''  -
   	2\xi \Bigl (\frac{A'}{A} + \frac{2r'}{r}\Bigr )\phi \phi'
\nnn \inch
   	+ \Bigl (\frac 12  - 2\xi - 3\xi^2 \phi^2 \Psi \Bigr)\phi'^2
		\biggr],
\ear
  and $G^i_k [\{\}]$ is the Einstein tensor of Riemannian space.
  The stress-energy tensor of the scalar field $\psi $ is
\beq\label{17}
   	T^k_i [\psi] = \frac{\eta_2 }2 A\psi'^2 \diag (1, -1, 1, 1)
   		+ \delta^k_i W(\psi).
\eeq

  The scalar field equations and three independent combinations of the
  Einstein--Cartan equations read
\bearr                            \label{18}
   	(1 - 6\xi^2 \phi^2 \Psi )\frac{(Ar^2 \phi')'}{r^2 } -
   	\frac{6\eta_1 }{\kappa }\xi^2 \phi\phi'^2 \Psi^2 A
\nnn \ \ \
	- \eta_1 \frac{dV}{d\phi }
   	+ \xi \phi \biggl [ A'' + \frac{4Ar''}{r} +\frac{4A'r'}{r}
\nnn \inch
	+ \frac{2Ar'^2 }{r^2 } - \frac2 {r^2 }\biggr] = 0,
\yyy 		\label{19}
   	(Ar^2 \psi')' = \eta_2 r^2 dW/d\psi \ ,
\yyy
   	(A'r^2 )' = - 2\eta_1 r^2 (V + W)\Psi
\nnn \ \ \		\label{20}
   	+ 2\xi r^2 A\Psi \biggl[\phi \phi'' + \phi'^2  +
   	2\Bigl (\frac{A'}{A} + \frac{r'}{r}\Bigr )\phi\phi' \biggr],
\yyy                       \label{21}
   	2\frac{r''}{r} = - \eta_1 \eta_2 \psi'^2 \Psi
    	+ \Psi \Big[ 2\xi \phi \phi''
\nnn \cm
	+ (- 1 + 2\xi + 6\xi^2 \phi^2 \Psi )\phi'^2 \Big],
\yyy		\label{22}
   	A(r^2)'' - r^2 A'' = 2 + 2\xi r^2 \phi\phi'\Psi \Bigl(
		\frac{2Ar'}{r} - A' \Bigr).
\ear
  It should be noted that the scalar field equation (\ref{18})
  follows from (\ref{19})--(\ref{22}).

  Noteworthy, in the Einstein-Cartan equations (\ref{18})--(\ref{22}) the
  terms induced by torsion contain the factor $\xi^2$, i.e., they exist due
  to nonminimal coupling of the $\phi$ field with space-time curvature. 
  It is this nonminimal coupling
  that makes ECT solutions different from those of GR, and in particular,
  as we shall see, it makes possible to obtain wormhole 
  solutions with non-phantom scalar fields $\phi$ and $\psi$ (that is, 
  with $\eta_1 = \eta_2 = 1$).

\section{Exact solution}

  Let us now try to find an exact solution to the Einstein--Cartan
  equations with canonical (normal) scalar fields: $\eta_1 = \eta_2 = +1$.
  Since finding solutions with given nonzero scalar field potentials is a
  hard problem even in the simpler case of GR, let us use the inverse
  problem method, by analogy with \cite{BF06, BBS12}, that is, consider the
  potentials as unknowns to be found but specify some other functions in
  accord with the desired properties of the solution. Since there are two
  potentials $V(\phi)$ and $W(\psi)$, we can choose two such functions.

  First, let us choose the function $r(u)$ in a way suitable for a wormhole
  solution:
\beq\label{23}
   	r(u) = \sqrt{u^2  + b^2 } = b^2 \sqrt{x^2  + 1},
\eeq
  where $x = u/b$, and $b$ is an arbitrary constant (the throat radius).
  Evidently, $r' = 0$ and $r'' > 0$ at $u=0$, as required.

  Second, we take the expression for $\phi (x)$ in the form
\beq\label{24}
   	\phi (x) = (\kappa \xi (x^2 + 1))^{- 1/2} \ ,
\eeq
  then \eq (\ref{21}) gives
\bearr    \label{25}
   	\frac2 {b^2 (x^2  + 1)^2 } = - \frac{\kappa (x^2+ 1)}{x^2}\psi'^2
	+ \frac{6\xi - 1}{\xi b^2 (x^2  + 1)^2 }
\nnn \inch
   	+ \frac{4}{b^2 x^2 (x^2  + 1)^2 }.
\ear
  \eq (\ref {25}) is solved for $\xi = 1/4$ and
\beq		\label{26}
   	  \psi'^2  = \frac{4}{\kappa b^2 (x^2  + 1)^3}.
\eeq
  whence
\beq		\label{27}
   	\psi (x) = \psi_{0} \pm 2x [\kappa (x^2  + 1)]^{- 1/2}.
\eeq

  Now, taking into account (\ref {23}) and (\ref {24}), we can obtain
  the metric function $A(x)$ from (\ref {22}). It is convenient to do that
  by bringing (\ref{22}) to the form
\[
	 B'' + 2B' \frac{2x^2 +1}{x(x^2 +1)} + \frac{2}{(x^2 +1)^2} =0
\]
  in the dimensionless variables $x$ and $B = A/(x^2{+}1)$; the prime means
  here $d/dx$. As a result, we obtain
\bearr             \label{28}
   	A(x) = (x^2 + 1)\Bigl [B_0 - \frac{C_1}{x} - C_1 \arctan x
\nnn \cm
   	- 2\frac{\arctan x}{x} - \arctan^2 x \Bigr ],
\ear
  where $C_1$ and $B_0$ are integration constants. Since $A(x)$
  should be regular at all $x$ including $x=0$, we put $C_1 = 0$.
  It means that our family of regular solutions contains only metrics
  symmetric with respect to the throat $x=0$.

  The asymptotic expression for $A(x)$ as $x\to \pm \infty $ can be
  presented as
\beq\label{29}
   	A(x) \simeq \Bigl (B_0 - \frac{\pi^2 }{4}\Bigr )x^2  + 1 + B_0
   			- \frac{\pi^2 }{4} \mp \frac{\pi }{3x}.
\eeq
  Thus the following cases are distinguished:
\begin{description}
\item[(i)]
	$B_0 > \pi^2/4$. The solution describes a traversable
	wormholes with two AdS asymptotic regions (AdS-AdS);
\item[(ii)]
	$B_0 = \pi^2/4$. The solution describes a twice asymptotically flat
	(M--M) traversable wormhole.
\item[(iii)]
	$B_0  < \pi^2/4$. We obtain configurations with two de Sitter
	asymptotics (dS--dS), which contain static regions only if $B_0 >
	2$. If $B_0 < 2$, the solution describes a pure KS cosmology
	interpolating between two dS states; if $B_0 = 2$, then two KS
	epochs are separated by a double horizon.
\end{description}

  Plots of $B(x) = A(x)/(x^2  + 1)$ with different $B_0$ are
  shown in Fig.\,1.
\begin{figure}
\centering
\includegraphics[width=7.5cm,height=5.5cm]{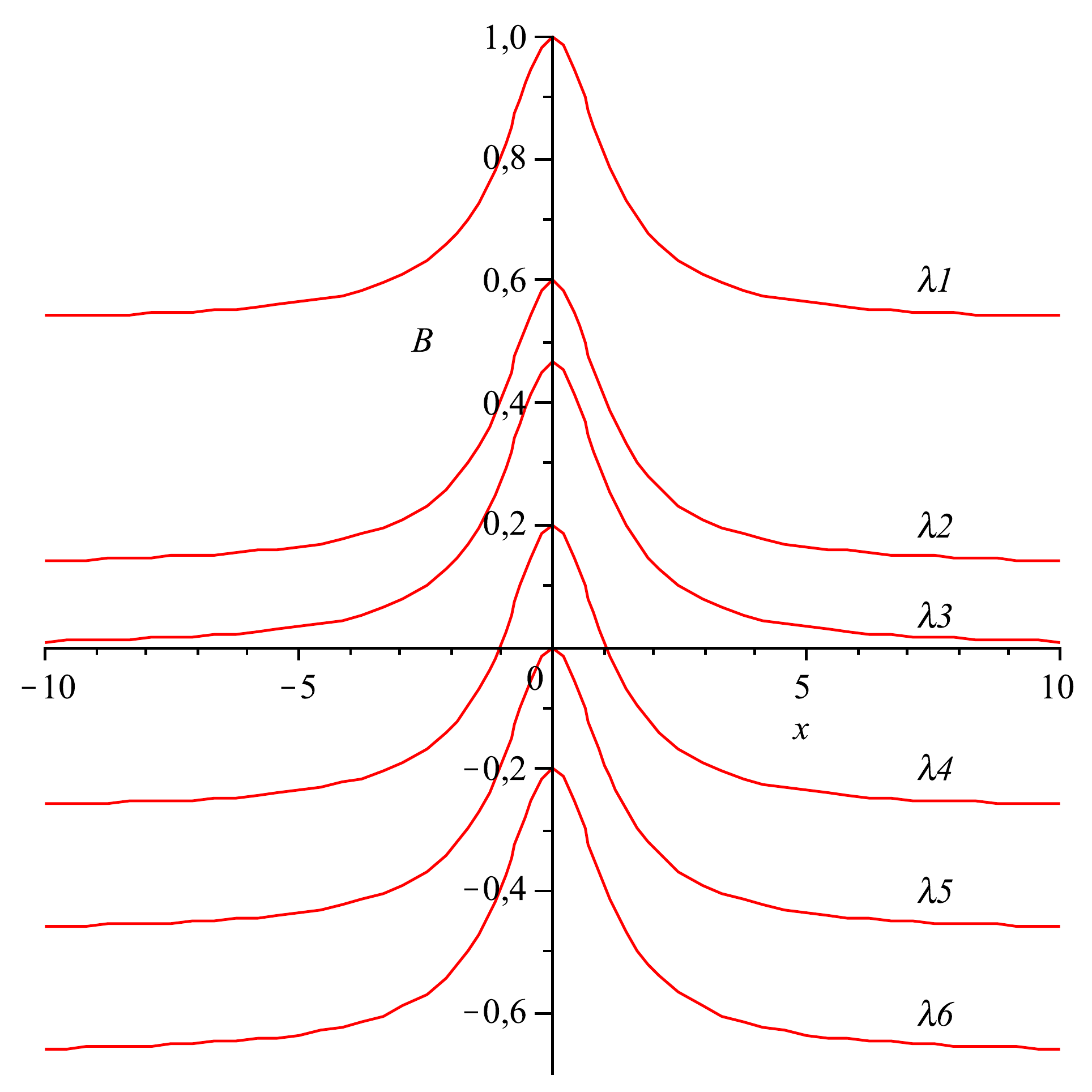}
\caption{\small
	 Plots of $B(x)$. Curves $\lambda 1$--$\lambda 6$ correspond to
	 $B_0 = 3,\ 2.6,\ \pi^2/4,\ 2.2,\ 2.0,\ 1.8$, }
\bigskip
\end{figure}

  In the twice asymptotically flat case (ii), comparing (\ref{29}) with the
  Schwarzschild expression $A = 1 - 2Gm/r$ and taking into account that
  $r \approx u = bx$ at large $x$, we find the Schwarzschild mass $m$:
\beq                            \label{mass}
      m = \frac{\pi b}{6G} = \frac{\pi}{6} m_{\rm pl} \frac{b}{l_{\rm pl}},
\eeq
  where $m_{\rm pl} = 1/\sqrt{G}$ and $l_{\rm pl} = \sqrt{G}$ are the Planck
  mass and length, respectively. It is easy to estimate that if we suppose
  that the wormhole is large enough for transportation purposes, say, $b=10$
  m, then the mass will be of the order $1.5 \ten{31}$ g, which makes about
  1/130 of the Sun's mass. The gravitational field in such a wormhole will
  be evidently too strong for a human being to survive.

  Let us find other quantities characterizing the solution. Choosing in
  (\ref {27}), without loss of generality, the plus sign and $\psi_0 = 0$,
  we obtain for $W(x)$ from (\ref {19}):
\bearr     \label{30}
   	W(x) = W_{0} + (8\kappa b^2 (x^2+ 1)^2 )^{- 1}\biggl[
	8B_0 x^2 (2 + x^2)
\nnn
   	+ 25x^2 (5 {+} 4x^2) - 2(75x^4 {+} 125x^2 {+} 32)\frac{\arctan x}{x}
\nnn\cm
   	- (75x^{4} + 150x^2  + 67)\arctan^2 x \biggr],
\ear
  where $W_{0}$ is an integration constant. Like $B(x)$, the function $W(x)$
  is even. Plots of $W(x)/8\kappa b^2 \to W(x)$ with $W_0 = 0$ are shown in
  Fig.\,2. Note that, in accord with the wormhole symmetry, the function
  $W(x)$ is even.

\begin{figure}
\centering
\includegraphics[width=7.5cm,height=5.5cm]{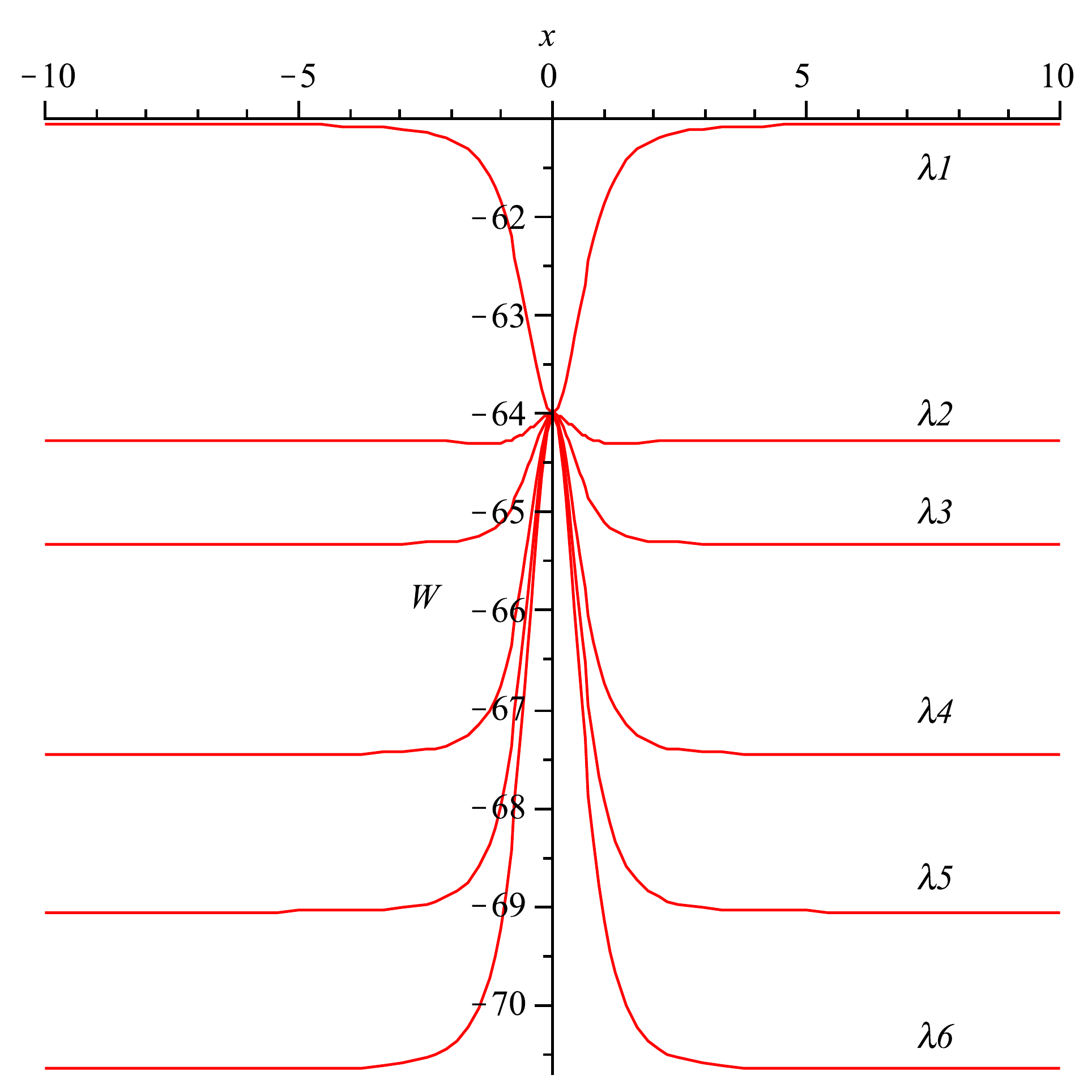}
\caption{\small
	Plots of $W(x)$. The curves are marked in the same way as in
	Fig.\,1.}
\bigskip
\end{figure}

  In terms of $\psi$ the potential $W$ is
\bearr     \label{31}
   	W(\psi) = W_0 + \frac{1}{128\kappa b^2}\biggl[
		8B_0 \psi^2 (8 - \psi^2 ) - 25\psi^4
\nnn \
   	+ 500\psi^2  + 8(\psi^{4} - 8\psi^2  - 134)\arctan^2 \psi
\nnn
   	+ 4(9\psi^4 {-} 122\psi^2 {-} 256)
   	\sqrt{4 {-} \psi^2}\, \frac{\arctan \psi}{\psi}\biggr],
\ear
  where $\psi$ stands for $\sqrt{\kappa}\psi $.

  From equation (\ref {20}) one derives $V(x)$:
\bearr     \label{32}
   	V(x) = \frac{1}{8\kappa b^2 (x^2  + 1)^2} \Bigl [- 8\kappa b^2 W_0
\nnn\
   	- 8B_0 (4x^4 + 6x^2  + 1) -100x^{4} - 101x^2  + 16
\nnn\
   	+ 2(99x^{4} + 149x^2  + 32)\frac{\arctan x}{x}
\nnn\
   	+ (99x^{4} + 182x^2  + 75)\arctan^2 x \Bigr ].
\ear
  Plots of $V(x)/8\kappa b^2  \to V(x)$ with $W_{0} = 0$ are
  shown in Fig.\,3.
\begin{figure}
\centering
	\includegraphics[width=7.5cm,height=5.5cm]{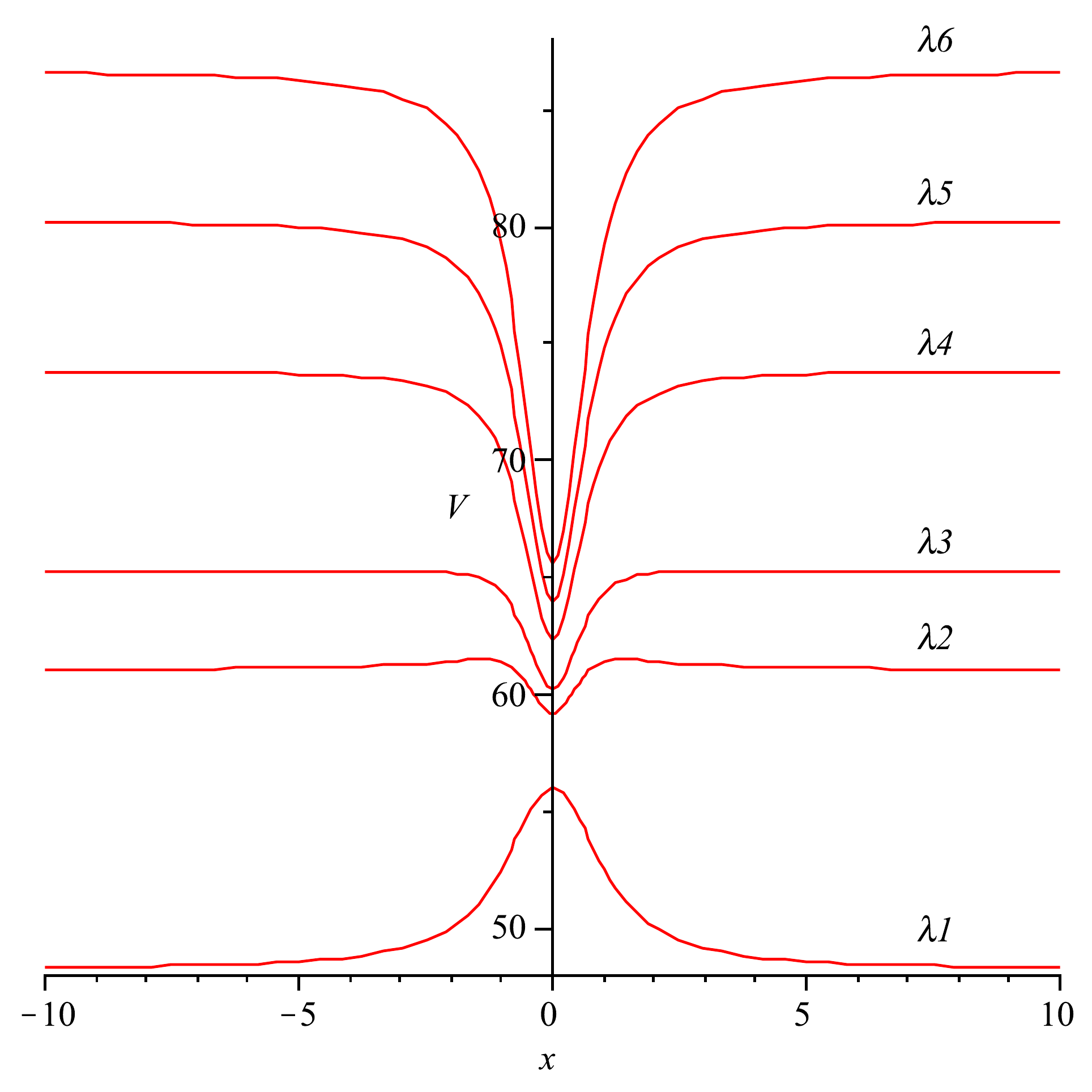}
\caption{\small
	Plots of $V(x)$. The curves are marked in the same way as in
	Fig.\,1.}
\bigskip
\end{figure}

  In terms of $\phi $ we obtain
\bearr \label{33}
   	V(\phi) = \frac{1}{8\kappa b^2} \Bigl [- 8\kappa b^2 W_{0}
\nnn \ \ \
   	+ 8B_0 (\chi^{4} + 2\chi^2  - 4)
   	+ 17\chi^{4}+ 99\chi^2  - 100
\nnn \ \ \
   	- 2(18\chi^4 {+} 49\chi^2 {-} 99)\frac{\arctan U}{U}
\nnn  \ \ \
   	- (8\chi^4 + 16\chi^2  - 99)\arctan^2 U \Bigr ] .
\ear
  where $\chi = \sqrt{\kappa \xi}\phi$, and $U = \sqrt{1 -\chi^2}/\chi$.

  The expression for the squared trace of the torsion $S^2  = S_k S^k $
  has the form
\beq\label{34}
  	S^2  = - \frac{9A(x)}{b^2 x^2 (x^2  + 1)^2 } .
\eeq
  Thus $S^{i}$ is a spacelike vector for $A > 0$ and a timelike one for
  $A < 0$, i.e., in a KS cosmology. The torsion turns out to be singular at
  the throat $x=0$. On the other hand, at large $x$ it decays by the laws
\bearr             \label{37}
   	S^2 \Big|_{x\to \pm \infty }\sim x^{-4} \to 0\qquad
   				\mbox{for}\ B_0 \neq \pi^2 /4,
\nnn
   	S^2 \Big|_{x\to \pm \infty }\sim x^{- 6} \to 0\qquad
   				\mbox{for}\ B_0  = \pi^2 /4 .
\ear

\section{Concluding remarks}

  We have found a family of exact static, spherically symmetric solutions
  with non-phantom scalar fields, describing asymptotically flat or AdS
  wormholes. Its existence proves that not only wormhole throats but also
  wormholes as global configurations are possible in the ECT due to nonzero
  space-time torsion, without NEC or WEC violation. In our view, it is a
  result of interest despite the special nature of this solution
  (in particular, a special value of the nonminimal coupling constant,
  $\xi = 1/4$) and some unpleasant physical properties of such wormholes:
  see the relation (\ref{mass}) between their Schwarzschild masses and
  throat radii, making the enormous wormhole gravity so unfavorable for
  life. Let us recall for comparison that in GR with a phantom
  scalar field a wormhole may have zero mass combined with a throat of
  arbitrary size \cite{br73,h_ellis}. Such solutions are ``force-free'',
  or ultrastatic, in the sense that $g_{00} \equiv 1$ in the whole space.
  On the other hand, if we try to substitute $A = g_{00} \equiv 1$
  to \eqs (\ref{18})--(\ref{22}) using the ansatz (\ref{23}), it follows
  from (\ref{22}) that either $\xi =0$ or $\phi = \const$, which means
  that there is no torsion [see (\ref{4})], and wormhole solutions are
  impossible. Though, one cannot exclude their existence with $r(x)$ other
  than (\ref{23}).

  Another problem with the above family of solutions is the divergent torsion
  at the throat $x=0$, although the metric, the scalar fields and their
  potentials are perfectly regular in the whole space-time. One can show,
  however, that this singularity is not a generic property of such solutions
  but is an artifact of the method used for solving the equations. Indeed,
  the singularity in (\ref{34}) is obtained due to the ansatz (\ref{24})
  for the scalar $\phi$, which leads to $\Psi = \kappa (x^2+1)/x^2$.
  Quite evidently, choosing a slightly different $\phi$ (let us say,
  $\phi = [\kappa\xi (x^2 + a^2)]^{-1/2}$ with $a$ slightly smaller than 1)
  will make the quantities $\Psi$ and $S^2$ finite and regular at all $x$,
  preserving the qualitative properties of our solution, but the results
  would not take such a simple and observable form.


\small
\begin{thebibliography}{99}

\bibitem{Pad03}                            
   	T. Padmanabham, Phys. Rep.  {\bf {380}}, 335  (2003).

\bibitem{PR03}                               
	P.J.E. Peebles and B. Ratra, Rev. Mod. Phys.  {\bf {75}},
	599 (2003).

\bibitem{SS06}                                 
	V. Sahni and A.A. Starobinsky, Int. J. Mod. Phys. D  {\bf
	   {15}}, 2105 (2006).

\bibitem{LLWW11}                                 
	M. Li, X.-D. Li, S. Wang, and Y. Wand, Commun. Theor. Phys.
	{\bf {56}}, 525  (2011); astro-ph/1103.5870[CO].

\bibitem{G07}     
	A.M. Galiakhmetov, Grav. Cosmol. {\bf {13}}, 217 (2007).

\bibitem{G10}        
	A.M. Galiakhmetov, Class. Quantum Grav.  {\bf {27}}, 055008 (2010).

\bibitem{G11}           
	A.M. Galiakhmetov, Class. Quantum Grav.  {\bf {28}}, 105013 (2011).

\bibitem{Cartan23}         
	E. Cartan, Ann. Ec. Norm. Suppl.  {\bf {40}}, 325 (1923).

\bibitem{Cartan24}            
	E. Cartan, Ann. Ec. Norm. Suppl.  {\bf {41}}, 1 (1924).

\bibitem{Cartan25}               
	E. Cartan,  Ann. Ec. Norm. Suppl.  {\bf {42}}, 17 (1925).

\bibitem{HO07}                      
	F.W. Hehl and Yu.N. Obu\-khov, Ann. Fond. Louis Broglie
	{\bf 32}, 157 (2007); gr-qc/0711.1535.

\bibitem{HHKN76}                       
	F.W. Hehl, P. von der Heyde, G.D. Kerlik, and J.M. Nester,
	Rev. Mod. Phys.  {\bf {48}}, 393 (1976).

\bibitem{PBO}                             
	V.N. Ponomarev, A.O. Barvinsky, and Yu.N. Obukhov,
	{\it Geometrodynamics Methods and Gauge  Approach in the Theory of
	Gravity} (Energoatomizdat, Moscow, 1985, in Russian).

\bibitem{MG06}                               
	A.V. Minkevich and A.S. Garkun, Class. Quantum Grav.  {\bf {23}},
	4237 (2006); gr-qc/0512130.

\bibitem{Trautman}  
	A. Trautman, Einstein-Cartan theory, in: {\it Encyclopedia of
	Mathematical Physics}, Eds. J.-P. Fran\c coise, G.L. Naber, and
	S.T. Tsou (Elsevier, Oxford, 2006), p. 189.

\bibitem{BH11}         
	P. Baekler and F.W. Hehl, Class. Quantum Grav.  {\bf {28}}, 215017
	    (2011).

\bibitem{Van87}           
	M.A.J. Vandyck, Class. Quantum Grav. {\bf 4}, 683 (1987).

\bibitem{Odi89}              
	S.D. Odintsov, Europhys. Lett. {\bf {8}}, 309 (1989).

\bibitem{BO89}                  
	I.L. Buchbinder and  S.D. Odintsov, Europhys. Lett. {\bf 8}, 595 (1989).

\bibitem{Kal81}                    
	M.W. Kalinowski, Acta phys. austr. {\bf {23}}, 641 (1981).

\bibitem{Ger82}                       
	G. German, Class. Quantum Grav. {\bf 2 }, 455 (1982).

\bibitem{VP88}                           
	Yu.S. Vladimirov and  A.D. Popov, Vestnik Mosk. Univ., Fiz.,
	Astron.  {\bf {4}}, 28 (1988).

\bibitem{AKR88}                             
	K. Akdeniz, A. Kizilers\"u, and E. Rizaoglu, Phys. Lett. B
	{\bf 215}, 81 (1988).

\bibitem{Kuu89}                                
	P. Kuusk, Gen. Relativ. Gravit. {\bf {21}}, 185 (1989).

\bibitem{Bak90}   
	W.M. Baker, Class. Quantum Grav. {\bf {7}}, 717 (1990).

\bibitem{CCSV07}     
   	S. Capozziello, R. Cianci, C. Stornaiolo and S. Vignolo,
   	Class. Quantum Grav. {\bf {24}}, 6417 (2007); gr-qc/0708.3038.

\bibitem{CCSV08}        
   	S. Capozziello, R. Cianci, C. Stornaiolo and S. Vignolo,
   	Phys. Scr. {\bf {78}}, 0605010 (2008); 	gr-qc/0810.2549.

\bibitem{hoh-vis}
	D. Hochberg and M. Visser,
	Phys. Rev. D {\bf 56}, 4745 (1997); gr-qc/9704082.

\bibitem{vis-book}
	M. Visser, {\it Lorentzian Wormholes: from Einstein to Hawking}
	(AIP, Woodbury, 1995).

\bibitem{lobo-rev}
	F.S.N. Lobo, Exotic solutions in general relativity:
	traversable wormholes and ``warp drive'' spacetimes. Arxiv: 0710.4474.

\bibitem{BR-book}
	K.A. Bronnikov and S.G. Rubin, {\it Black Holes, Cosmology and Extra
	Dimensions} (World Scientic, 2012).

\bibitem{br73}
        K.A. Bronnikov, 
        Acta Phys. Polon., B4, 251 (1973).

\bibitem{h_ellis}
	H. Ellis, J. Math. Phys. {\bf 14}, 104 (1973)

\bibitem{BF06}                   
	K.A. Bronnikov and J.C. Fabris, Phys. Rev. Lett. {\bf 96}, 251101
        (2006); gr-qc/0511109.

\bibitem{BBS12}                     
        S.V. Bolokhov, K.A. Bronnikov, and M.V. Skvortsova, Class. Quantum Grav.
        {\bf 29}, 245006 (2012).

\bibitem{BVis99}                       
        C. Barcel\'o and M. Visser, Phys. Lett. B.  {\bf {466}}, 127 (1999);
        gr-qc/09908029.

\bibitem{BVis00}
	C. Barcel\'o and M. Visser, Class. Quantum Grav. {\bf 17}, 3843 (2000).

\bibitem{br-96}
	K.A. Bronnikov, 
        Grav. Cosmol. {\bf 2}, 221--226 (1996).

\bibitem{br-gr02}
	K.A. Bronnikov and S. Grinyok,
	Festschrift in honour of Prof. Mario Novello., Rio de Janeiro, 2002;
	gr-qc/0205131.

\bibitem{br-JMP}
	K.A. Bronnikov, 
        \JMP {43} 6096--6115 (2002); gr-qc/0204001.

\bibitem{br-star07}
	K.A. Bronnikov and A.A. Starobinsky,
        Pis'ma v ZhETF {\bf 85}, 1, 3-8 (2007);
        JETP Lett. {\bf 85}, 1, 1-5 (2007); gr-qc/0612032.

\bibitem{br-Pol}
	K.A. Bronnikov, 
        Acta Phys. Polon.{\bf B 32}, 3571--3592 (2001); gr-qc/0110125.

\bibitem{br-gr01}
	K.A. Bronnikov and S.V. Grinyok,
        \GC {7} 297--300 (2001); gr-qc/0201083.

\bibitem{br-gr04}
	K.A. Bronnikov and S.V. Grinyok,
	\GC {10} 237--244 (2004); gr-qc/0411063.

\bibitem{BSS10}
	K.A. Bronnikov, M.V. Skvortsova and A.A. Starobinsky,
        \GC {16} 216--222 (2010); ArXiv: 1005.3262.

\bibitem{KS97}           
	V.G. Krechet and D.V. Sadovnikov, Grav. Cosmol. {\bf 3}, 133 (1997).

\bibitem{krechet1}
	V.G. Krechet and D.V. Sadovnikov, Grav. Cosmol. {\bf 13}, 269 (2007).

\bibitem{krechet2}
	V.G. Krechet and D.V. Sadovnikov, Grav. Cosmol. {\bf 15}, 337 (2009).

\bibitem{BKS12}
	K.A. Bronnikov,	V.G. Krechet, and Jos\'e P.S. Lemos,
     	\PRD {87} 084060 (2013); ArXiv: 1303.2993. 

\end{thebibliography}
\end{document}